\pdfoutput=1

\documentclass[reprint,amsmath,amssymb,floatfix,aps,longbibliography]{revtex4-1}
\usepackage{graphics,graphicx,float}

\usepackage[pdftex,
            pdfauthor={Steven A. Frank},
            pdftitle={},
            pdfsubject={Natural selection. VI. Partitioning the information in fitness and characters by path analysis}]{hyperref} 

\def\citeyear{\citep}
\def\autocite{\citep}

\newcommand{\zbar}{\bar{z}}
\newcommand{\wbar}{\bar{w}}
\newcommand{\mbar}{\bar{m}}
\newcommand{\gbar}{\bar{g}}
\newcommand{\cov}{{\hbox{\rm Cov}}}

\newcommand{\KL}[2]{\mathcal D\left(#1||#2\right)}

\newcommand{\J}{J}

\newcommand{\qfraclog}{\log\left(\frac{q_i'}{q_i}\right)}

\newcommand{\GDs}{\GD_s}
\newcommand{\GDn}{\GD_n}
\newcommand{\GDt}{\GD_t}
\newcommand{\GDc}{\GD_c}
\newcommand{\GDh}{\GD_h}

\newcommand{\preg}[2]{\Gb_{#1\cdot#2}}

\newcommand{\Ga}{\alpha}
\newcommand{\Gb}{\beta}
\newcommand{\Gd}{\delta}
\newcommand{\GD}{\Delta}
\newcommand{\Ge}{\epsilon}

\newcommand{\Gf}{\phi}

\DeclareMathOperator{\E}{E}
\newcommand{\dd}{{\hbox{\rm d}}}

\newcommand{\Eq}[1]{eqn~\ref{eq:#1}}

\newcommand{\boldrule}{\hrule height 1.2pt}
\newcommand{\noterule}{\medskip\boldrule\medskip}	% for notes

\newcommand{\noterulenote}[1]{\null}

\newcount\BoxNum \BoxNum 1\relax
\makeatletter
\newcommand{\boxlabel}[1]{%
  \protected@write \@auxout {}{\string \newlabel {box:#1}{{\the\BoxNum}{\thepage}{\noexpand\relax}%
  	{\@ifundefined{hyper@@anchor}{\relax}{box.\the\BoxNum}}%
  	{}}}%
  \@ifundefined{hyper@@anchor}{}{\hypertarget{box.\the\BoxNum}{}}%
  \advance\BoxNum 1\relax}
\makeatother
\newcommand{\Boxx}[1]{Box~\ref{box:#1}}
\newcommand{\BoxLabel}{Box~\the\BoxNum}

\begin{document}

\title{Natural selection. VI. Partitioning the information in fitness and characters\\ by path analysis}

\author{Steven A.\ Frank}
\email[email: ]{safrank@uci.edu}
\homepage[homepage: ]{http://stevefrank.org}
\affiliation{Department of Ecology and Evolutionary Biology, University of California, Irvine, CA 92697--2525  USA}

\begin{abstract}

Three steps aid in the analysis of selection.  First, describe phenotypes by their component causes. Components include genes, maternal effects, symbionts, and any other predictors of phenotype that are of interest.  Second, describe fitness by its component causes, such as an individual's phenotype, its neighbors' phenotypes, resource availability, and so on.  Third, put the predictors of phenotype and fitness into an exact equation for evolutionary change, providing a complete expression of selection and other evolutionary processes.  The complete expression separates the distinct causal roles of the various hypothesized components of phenotypes and fitness.  Traditionally, those components are given by the covariance, variance, and regression terms of evolutionary models.  I show how to interpret those statistical expressions with respect to information theory.  The resulting interpretation allows one to read the fundamental equations of selection and evolution as sentences that express how various causes lead to the accumulation of information by selection and the decay of information by other evolutionary processes.  The interpretation in terms of information leads to a deeper understanding of selection and heritability, and a clearer sense of how to formulate causal hypotheses about evolutionary process.  Kin selection appears as a particular type of causal analysis that partitions social effects into meaningful components\footnote{\href{http://dx.doi.org/10.1111/jeb.12066}{doi:\ 10.1111/jeb.12066} in \textit{J. Evol. Biol.}}\footnote{Part of the Topics in Natural Selection series. See \Boxx{preface}.}. 

\end{abstract}

\maketitle

\begin{quote}
The path method $\ldots$ is not so much concerned with prediction as [it is with] the proposal of a \textit{plausible interpretation} of the relationships between the variables. In other words, path analysis is concerned with erecting a causal structure compatible with the observed data \autocite[p.~3]{li75path}.
\end{quote}

\section*{Introduction}

Populations accumulate information by natural selection.  The amount of information may be expressed by classical information theory \autocite{frank12naturalc}.  That purely informational expression describes phenotypes and fitness abstractly, without consideration of the explicit causes that determine phenotypic traits and their association with fitness. Here, I partition phenotypes and fitness into their component causes.  

For phenotypes, we must track the influence of genes, symbionts, maternal effects and other potential causes.  The components of phenotype lead to explicit models of character expression and heritability.  For fitness, we must track how different characters and external forces combine to determine success.  An individual's fitness may, for example, depend on a combination of its own phenotype and the phenotypes of its neighbors.  

I put those explicit causal components of phenotype and fitness into the fundamental expressions of selection and evolutionary change.  I recover an expanded concept of heritability, a precise understanding of Fisher's fundamental theorem, and a general form of the equations of selection for multiple characters.  With those tools, the following article clarifies kin selection and other social processes \autocite{frank13naturalvii}.  

I presented much of this material in \textcite{frank97the-price,frank98foundations}.  Here, I pursue four goals.  First, I express the key partitions of phenotypes and fitness with respect to my new information theory interpretation of selection \autocite{frank12naturalc}.  Second, the information expressions translate the traditional regression and variance terms of selection into more meaningful descriptions of cause and consequence.  Third, the partitions of phenotype and fitness provide the basis for replacing outdated concepts of kin selection with a solid conceptual foundation \autocite{frank13naturalvii}.  Fourth, I emphasize simplicity, presenting the mathematical material at the most basic level consistent with the concepts. The original publications contain more detail \autocite{frank97the-price,frank98foundations}.

Mathematically, little is required beyond simple forms of statistical regression and the location of points in coordinate systems.  Although I use only basic mathematics, the article is nonetheless challenging.  I cover a wide array of problems at a very general level, with emphasis on the connections between seemingly different topics.  That sustained abstraction and synthesis provide both significant rewards and demanding challenges. 

It may seem that the basic problems of selection and kin interactions were solved long ago.  Why do we need to revisit those topics?  In fact, our understanding of natural selection and kin selection has continued to advance over the past few decades.  Those advances have developed while the old formulations have remained.  The core of the subject has become cluttered with incompatible expressions from different eras, derived in different contexts. One can no longer go forward without first resetting the foundations.  

\begin{figure}[H]
\begin{minipage}{\hsize}
\parindent=15pt
\noterule
{\bf \noindent\BoxLabel. Topics in the theory of natural selection}
\noterule
\noindent This article is part of a series on natural selection.  Although the theory of natural selection is simple, it remains endlessly contentious and difficult to apply.  My goal is to make more accessible the concepts that are so important, yet either mostly unknown or widely misunderstood.  I write in a nontechnical style, showing the key equations and results rather than providing full derivations or discussions of mathematical problems.  Boxes list technical issues and brief summaries of the literature.      
\noterule
\end{minipage}
\end{figure}
\boxlabel{preface}

\section*{Selection}

I briefly review the general equations for selection and evolution.  Recent articles in this series provide full details \autocite{frank12naturalb,frank12naturalc}.  

\subsection*{The Price equation}

Consider an initial population.  Let $\zbar$ be the average in the population of some value (phenotype).  A second population has average value $\zbar'$.  Total change between the populations is $\GD\zbar = \zbar'-\zbar$.  Split the total change into two components
\begin{equation}\label{eq:totalSym}
  \GD\zbar = \GDs\zbar + \GDc\zbar.
\end{equation}
The first term, $\GDs$, is the part of the total change caused by selection.  The second term, $\GDc$, is the remaining part of total change by all other causes. 

To evaluate these terms, we write the average value as $\zbar=\sum q_iz_i$. The index $i$ divides the population in any way that we choose.  We may use $i$ to label by different individuals, by different groups, by genotype, or by any other partition of the population.  The frequency of a type $i$ in the population is $q_i$.  The phenotype associated with $i$ is $z_i$.  The average value in the second population is $\zbar'=\sum q_i'z_i'$.

We define selection as changes in frequency, holding constant phenotype
\begin{equation*}
  \GDs\zbar = \sum q_i'z_i - \sum q_iz_i.
\end{equation*}
Here, the populations differ in their frequencies, $\GD q_i = q_i' - q_i$, but we have held the phenotype values constant at $z_i$ in both populations.  Using $\GD q_i$ for frequency change, we write
\begin{equation}\label{eq:selDef}
  \GDs\zbar = \sum \GD q_i z_i.
\end{equation}

To obtain the total change, we need the changes in phenotype holding constant the frequencies
\begin{equation*}
  \GDc\zbar = \sum q_i'z_i' - \sum q_i'z_i.
\end{equation*}

\begin{figure}[H]
\begin{minipage}{\hsize}
\parindent=15pt
\noterule
{\bf \noindent\BoxLabel. Price equation: difference of a product}
\noterule
\noindent The Price equation simply expands a difference into multiple terms.  Consider, for example, the difference of the product of $x$ and $y$, which we write as $\GD \left(xy\right) = x'y'-xy$.  We can expand the difference of the product as
\begin{equation*}
  \GD (xy) = \left(x + \GD x\right)  \left(y+\GD y\right)  - xy
\end{equation*}
which yields
\begin{equation*}
  \GD (xy) = \left(\GD x\right) y + x \left(\GD y\right) + \GD x \GD y.
\end{equation*}
This expression shows that the difference of a product is the difference of the first term holding the second term constant, plus the difference of the second term holding the first term constant, plus the product of the two differences.  

We can simplify the difference expansion by combining a pair of terms on the right-hand side.  Noting that $x'=x+\GD x$, we can combine the last two terms into one, yielding
\begin{equation*}
  \GD (xy) = \left(\GD x\right) y + x' \left(\GD y\right).
\end{equation*}
The derivation of the Price equation follows the rule for the difference of a product
\begin{align*}
  \GD\zbar &= \GD\sum q_iz_i \\
                  &=\sum\GD\left(q_iz_i\right) \\
                  &=\sum\left(\GD q_i\right)z_i + \sum q_i'\left(\GD z_i\right). 
\end{align*}
The value of the Price equation arises from identifying $\sum\left(\GD q_i\right)z_i$ as the part of total change caused by selection.  Selection acts on phenotype at a fixed point in time, so it makes sense to consider selection as the partial difference in frequency holding phenotype constant.  When we use log fitness for the phenotype, $m\equiv z$, we get an exact correspondence between the selection term and the increase in information expressed by classical information theory (\Eq{mJ}).  That correspondence supports interpreting $\sum\left(\GD q_i\right)z_i$ as selection.     
\noterule
\end{minipage}
\end{figure}
\boxlabel{price}

Here, the populations differ in their phenotype, $\GD z_i = z_i'-z_i$, but we have fixed the frequency at $q_i'$.  We use the final frequencies in the second population, $q'$, because they provide the proper reference for final phenotype after change (\Boxx{price}).  Using $\GD z_i$ for phenotypic changes, we write
\begin{equation*}
  \GDc\zbar = \sum q_i'\GD z_i.
\end{equation*}
The total change from \Eq{totalSym} can now be written (\Boxx{price}) as a form of the Price equation
\begin{equation}\label{eq:price}
  \GD\zbar = \sum \GD q_i z_i + \sum q_i'\GD z_i.
\end{equation}

\subsection*{Classical expressions of covariance, regression and variance}

The definition of fitness is
\begin{equation}\label{eq:fitDef}
  q_i' = q_i \frac{w_i}{\wbar},
\end{equation}
where $w_i$ is the fitness of type $i$, and $\wbar$ is average fitness.  The change in frequency is  
\begin{equation*}
  \GD q_i = q_i \left(\frac{w_i}{\wbar}-1\right).
\end{equation*}
Thus, the change caused by selection can be written as a covariance between fitness and phenotype
\begin{equation}\label{eq:cov}
  \sum \GD q_i z_i = \sum q_i \left(\frac{w_i}{\wbar}-1\right)z_i = \cov(w,z)/\wbar.
\end{equation}
We can rewrite a covariance as a product of a regression coefficient and a variance term
\begin{equation}\label{eq:fitnessReg}
  \GDs\zbar = \cov(w,z)/\wbar = \Gb_{zw}V_w/\wbar,
\end{equation}
where $\Gb_{zw}$ is the regression of phenotype, $z$, on fitness, $w$, and $V_w$ is the variance in fitness. Selection equations are often expressed with these covariance, regression and variance terms.  Classical population genetics expressions for change in gene frequency also have this form, in which we
let $\zbar = p$ be the frequency of a gene in a population.

\section*{Information}

\textcite{frank12naturalc} showed that selection can be expressed in terms of information theory.  I briefly review the key points in this section.

\subsection*{Fitness and the gain in encoded information}

Fitness, $w$, describes relative changes in frequency.  Logarithms provide the natural scaling for relative changes. Using the expression for fitness in \Eq{fitDef}, we write log fitness as 
\begin{equation*}
  m_i = \log(w_i) = \log(\wbar) + \qfraclog.
\end{equation*}
Using $z \equiv m$ in the expression for selection (\Eq{selDef}), we have
\begin{equation*}
  \GDs\mbar = \sum\GD q_i m_i = \sum \GD q_i \qfraclog.
\end{equation*}
The classic information theory expression for the change in encoded information between two populations with frequencies $q'$ and $q$ is
\begin{equation}\label{eq:KLdef}
  \KL{q'}{q} =\sum q_i'\qfraclog.
\end{equation}
With that definition, we have
\begin{equation*}
  \GDs\mbar = \KL{q'}{q} + \KL{q}{q'},
\end{equation*}
in which the right hand side is known as the Jeffreys information divergence, $\J$. Thus, we can write the fundamental expression for the accumulation of information by natural selection as
\begin{equation}\label{eq:mJ}
  \GDs\mbar = \J.
\end{equation}
Because $z$ in \Eq{fitnessReg} is just a placeholder for any character, we can use $m$ in place of $z$ in that equation, yielding
\begin{equation*}
  \GDs\mbar = \Gb_{mw}V_w/\wbar.
\end{equation*}
Thus, the information accumulated by natural selection, $\J$, is equivalently expressed in terms of the regression coefficient and the variance, 
\begin{equation}\label{eq:Jvar}
  \J=\Gb_{mw}V_w/\wbar.
\end{equation}

\subsection*{Variance, regression and information}

The variance in fitness, $V_w$, is proportional to the information gain by natural selection, $\J$ (\Eq{Jvar}).  It is easy to understand why selection may be expressed in terms of information.  Selection is, in essence, a process by which populations gain information about the environment.  But why should the variance arise as an alternative description of selection?  

The usual view is that selection acts on differences within the population.  The greater the differences, the larger the variance and the greater the opportunity for selection.  But why exactly is the variance the correct measure of differences within the population, rather than some other measure of variation?  

Consider the definition of fitness (\Eq{fitDef}) given earlier
\begin{equation*}
  \frac{w_i}{\wbar} = \frac{q_i'}{q_i},
\end{equation*}
in which the relative fitness is the ratio of frequencies between the new and old population.  Relative fitness is, in essence, a measure of the separation between the new population and the old population, a comparison of $q'$ versus $q$.  Because the frequencies in each population must add to one, each separation between a pair $q_i'$ and $q_i$ must be balanced by opposite separations in other pairs.  

Thus, the variation in the $q_i'/q_i$ ratios measures the total separation of the new population from the old population.  In particular, the variance in those ratios---the variance in fitness---is like a distance between the new population and the old population.  That distance-like measure has units in terms of the information gain \autocite{frank12naturalc}.  The variance in fitness expresses an informational distance, the amount of information gained by selection.  

Information gain is measured on the logarithmic scale of frequency changes (\Eq{KLdef}).  The regression coefficient, 

\begin{figure}[H]
\begin{minipage}{\hsize}
\parindent=15pt
\noterule
{\bf \noindent\BoxLabel. Regression}
\noterule
\noindent Simple regression is based on the equation for a line
\begin{equation*}
  z = a + \Gb y,
\end{equation*}
in which $z$ is the outcome of interest, $y$ is a variable that is used to predict $z$, the term $\Gb$ is the slope of the line relating $z$ to $y$, and $a$ is the intercept, which is the value of $z$ when $y=0$.  The simple regression model is usually written as
\begin{equation*}
  z_i = a + \Gb_{zy} y_i + \Gd_i,
\end{equation*}
in which the $i$ subscripts denote values associated with different observations, and $\Gd_i$ is the residual as described below.  In some applications, it is convenient to make the intercept $a$ disappear, which we achieve by $y_i = x_i -a/\Gb_{zy}$, which gives
\begin{equation*}
  z_i = \Gb_{zx} x_i + \Gd_i.
\end{equation*}
This expression is equivalent to the previous one.  The only change is that $x$ differs from $y$ by a constant value.  The second expression uses $\Gb_{zx}$ in place of $\Gb_{zy}$. Those terms have the same value, but I use the term with $x$ to emphasize that the relation is now between $z$ and $x$.   In any regression model, we can make a similar substitution in which we change $y$ by a constant factor to get an $x$ value that makes the intercept disappear. 

From the perspective of regression analysis, $\Gb_{zx}x$ provides a prediction of $z$ given $x$.  The difference between the actual value and the predicted value is the residual (error), $\Gd_i = z_i - \Gb_{zx}x$.  Two changes in notation provide a cleaner expression. Write the regression coefficient as $b = \Gb_{zx}$, and drop the $i$ subscript, yielding
\begin{equation*}
  z = b x + \Gd,
\end{equation*}
where the variables implicitly range over $i$.  

Regression has a natural asymmetry.  In prediction, the value of $z$ is the predicted value given the predictor, $x$.  In a causal interpretation, in the sense of path analysis (\Boxx{cause}), the effect $z$ depends on the cause, $x$.  One must keep this asymmetry in mind to interpret regression equations correctly.  Proper notation helps.  We may write
\begin{equation*}
  z|x = b x + \Gd,
\end{equation*}
which emphasizes that the outcome, $z$, depends on the given fixed value of $x$.  We read $z|x$ as ``$z$ given $x$.'' If we take the average of both sides
\begin{equation*}
  \E(z|x) = b x,
\end{equation*}
where $\E(z|x)$ is the expectation of $z$ given $x$, in which ``expectation'' means the average value. On the right side, $\Gd$ disappears because the regression coefficient, $b$, is chosen so that the average value of the residual is zero, $\bar{\Gd}=0$.    
\noterule
\end{minipage}
\end{figure}
\boxlabel{regression}

\noindent $\Gb_{mw}$, transforms fitness from the linear scale, $w$, to the log scale, $m$, yielding the key expression given earlier for the change in log fitness (information) caused by selection
\begin{equation*}
  \GDs\mbar = \J=\Gb_{mw}V_w/\wbar.
\end{equation*}
It is common to think of a regression coefficient as a linear prediction estimated from data.  That interpretation misleads with regard to understanding the fundamental equations of selection.  Instead, the regression coefficient describes the consequence for the change in average value when transforming from one scale to another scale (Boxes \ref{box:regression} and \ref{box:scale}).  The proper way to read $\Gb_{mw}$ is a change in scale from $w$ to $m$ when evaluating the averages $\wbar$ and $\mbar$.

\subsection*{Phenotype as a change in the scaling of information}

Selection causes populations to accumulate information.  The measure of information is related to log fitness.  In the analysis of selection, we often focus on phenotypes rather than fitness.  Here, I show that, with respect to selection, one can think of the phenotypic scale simply as an alternative scale on which to measure information.

Begin with the expression given earlier for the change in log fitness
\begin{equation*}
  \GDs\mbar = \Gb_{mw}V_w/\wbar.
\end{equation*}
The regression coefficient, $\Gb_{mw}$, changes scale from fitness, $w$, to log fitness, $m$.  If we divide by $\Gb_{mw}$, we obtain
\begin{equation*}
  \frac{\GDs\mbar}{\Gb_{mw}} = V_w/\wbar.
\end{equation*}
The factor $1/\Gb_{mw}$ reverses the scale change, transforming from the logarithmic scale, $m$, to the linear scale, $w$.

The change in phenotype from \Eq{fitnessReg} can be written as
\begin{equation*}
  \GDs\zbar = \Gb_{zw}V_w/\wbar.
\end{equation*}
The regression $\Gb_{zw}$ changes scale from fitness, $w$, to phenotype, $z$, and $1/\Gb_{zw}$ reverses the direction of the change in scale.  Thus
\begin{equation*}
  \frac{\GDs\zbar}{\Gb_{zw}} = V_w/\wbar = \frac{\GDs\mbar}{\Gb_{mw}}.
\end{equation*}
Because the information accumulated by natural selection is $\GDs\mbar=\J$, we have
\begin{equation*}
  \GDs\zbar = \left(\frac{\Gb_{zw}}{\Gb_{mw}}\right)\J.
\end{equation*}
This expression describes the change in phenotype by selection in relation to the information gain, $\J$, rescaled by the transformation from the scale of information, $m$, to the scale of phenotype, $z$.  We may describe the scaling between the gain in information, $\J$, and change in phenotype caused by selection, $\GDs\zbar$, as
\begin{equation}\label{eq:alphaDef}
  \Ga_z = \frac{\Gb_{zw}}{\Gb_{mw}}.
\end{equation}
Thus we can write the relation between the change in phenotype and the gain in information as 
\begin{equation}\label{eq:alphaJ}
  \GDs\zbar = \Ga_z\J.
\end{equation}

\begin{figure}[H]
\begin{minipage}{\hsize}
\parindent=15pt
\noterule
{\bf \noindent\BoxLabel. Change in scale}
\noterule
\noindent In the regression model (\Boxx{regression}) with subscripts used explicitly for labeling types
\begin{equation*}
  \E(z_i|x_i) = b x_i.
\end{equation*}
If we consider subscripts for two different types, $k$ and $i$, we can write $\E(z_k|x_k) = b x_k$ and $\E(z_i|x_i) = b x_i$.  Subtracting these two equations from each other gives
\begin{equation*}
  \E(z_k-z_i|x_k-x_i) = b(x_k - x_i).
\end{equation*}
Using $\GD$ to denote a change between the $k$ and $i$ values
\begin{equation*}
  \E(\GD z|\GD x) = b(\GD x),
\end{equation*}
which we can write equivalently as
\begin{equation*}
  b = \Gb_{zx} = \frac{\E(\GD z|\GD x)}{\GD x},
\end{equation*}
which we read as: ``the regression of $z$ on $x$ is the expected change in $z$ for a given change in $x$ divided by the change in $x$.''  From this expression, we see that a regression coefficient is the expected change in scale for one variable in relation to another variable.  One can also think of the regression coefficient as a sort of generalization of differentiation.  For situations in which we can consider $z$ and $x$ as continuous variables with an underlying functional relationship, $z(x)$, it will often be the case that, as the changes become small, $\GD z \rightarrow 0$ and $\GD x\rightarrow 0$ with $x$ confined to a small range of values, then the regression coefficient approaches the derivative, $\Gb_{zx} \rightarrow \dd z/\dd x$.

Finally, the variables $x$ and $\Gd$ are uncorrelated, so that $\cov(x,\Gd) = 0$.  Regression uses all of the available information in $x$ about $z$.  Thus, any left over deviations, $\Gd$, cannot contain information about $z$, which is reflected in the lack of correlation between those variables.

When we have multiple predictors, or causes, $x_j = x_1, x_2,\ldots,x_n$, then the regression equation is 
\begin{equation*}
  z = \sum_j b_j x_j + \Gd,
\end{equation*}
where each $b_j$ is the partial regression of $z$ on $x_j$, holding constant the other predictor values.  Suppose, for example, that we have two predictors, $x_1$ and $x_2$. For notational convenience, let $x \equiv x_1$ and $y \equiv x_2$, so that the regression equation is
\begin{equation*}
  z = b_x x + b_y y + \Gd.
\end{equation*}
If, as above, we take the difference between two $x$ values, holding $y$ constant, we obtain
\begin{equation*}
  b_x = \Gb_{zx\cdot y} = \frac{\E(\GD z|\GD x,y)}{\GD x},
\end{equation*}
which we read as: ``the regression of $z$ on $x$, holding $y$ constant, is the expected change in $z$ for a given change in $x$ and a fixed value of y, divided by the change in $x$.''  This expression gives the expected change in scale between $z$ and $x$ for a given value of $y$.  If $z$, $x$, and $y$ are continuous variables with an underlying functional relationship, $z(x,y)$, then for small changes confined to a small range of predictor values for $x$ and $y$, it will often be the case that the regression approaches the partial derivative $\Gb_{zx\cdot y} \rightarrow \partial z/\partial x$.

\noterule
\end{minipage}
\end{figure}
\boxlabel{scale}

\begin{figure}[H]
\begin{minipage}{\hsize}
\parindent=15pt
\noterule
{\bf \noindent Box 4 --- continued\vskip2pt}
\noterule
\noindent These properties of regression follow from least squares.  The squared distance between predicted and observed values is the sum of squares, $\sum \Gd_i^2$.  Minimizing that distance gives the least value for the sum of squares---the least squares.  All properties here follow from that minimization.  Further aspects of regression depend on other assumptions. For example, many tests of statistical significance assume that the residuals have a normal distribution.  Certain interpretations require that the observations be linearly related to the predictors.  I do not use those further aspects and therefore do not require any assumptions about linearity or the distribution of observations and residuals.      
\noterule
\end{minipage}
\end{figure}

\section*{Causes of phenotype}

This section partitions the causes of phenotype into components.  The next section connects the causes of phenotype to the capture and transmission of information.  The following section partitions fitness into components, dividing the gain in information by selection into different causes.  Boxes \ref{box:regression}--\ref{box:nonlinearity} provide background on regression. \Boxx{partition} provides citations to the literature.

\subsection*{Overview}

Heritability describes the expected similarity in phenotype between different individuals \autocite{falconer96introduction}.  For example, we may define the predictors of phenotype as the set of alleles in an individual, and the heritability as the part of similarity between ancestors and descendants ascribed to those alleles.  Because sex and recombination break up particular combinations of alleles, adding up the effects of each individual allelic predictor often provides a good estimate of the similarity between different relatives caused by genetics.  

Alternatively, we may expand the set of predictors to include certain nonlinear combinations of alleles.  For example, we may have a predictor for the presence of allele A, another for the presence of allele B, and a third for the presence of both alleles.  Certain expanded predictor sets may give a more accurate description of similarity between closely related ancestor-descendant pairs that are likely to share the allelic combinations, but may give a less accurate description when the allelic pairs tend to be broken up during transmission.

Here, I am primarily interested in the information that a population accumulates by selection, and how different processes may reduce or alter the transmission of accumulated information.  My expressions include the classic genetic measures as special cases.  But I do not emphasize the connection to traditional genetics---the genetic interpretations are discussed in every basic textbook of\break  

\begin{figure}[H]
\begin{minipage}{\hsize}
\parindent=15pt
\noterule
{\bf \noindent\BoxLabel. Causes and predictors}
\noterule
\begin{quote}
Since path analysis depends on structure, and structure in turn depends on the cause-and-effect relationship among the variables, we shall first say a few words about the way these terms will be used $\ldots$ There are a number of formal definitions as to what constitutes a cause and what an effect. For instance, one may think that a cause must be doing something to lead to something else (effect). While this is clearly  one type of cause-and-effect relationship, we shall not limit ourselves to that type only. Nor shall we enter into philosophical discussions about the nature of cause-and-effect.  We shall simply use the words ``cause'' and ``effect'' as statistical terms similar to independent and dependent variables, or [predictor variables and response variables] \autocite[p.~3]{li75path}.
\end{quote}

\noindent I analyze causes of phenotypes and causes of fitness.  Here, I briefly comment on the word ``cause.''  The above quote and the epigraph come from Li's book on \textit{Path Analysis}.  Li's point concerns the distinction between three levels of analysis.  First, true causality describes the relations between actual forces and actual effects.  Whether such things can ever be studied or known directly remains a philosophical problem beyond our scope.  

Second, at the other extreme, multiple regression analysis from classical statistical theory concerns only correlations and variances. The standard theory explicitly disavows causal interpretation---correlation is not causation.  Regression arises by minimizing the distance between predicted outcomes and actual outcomes---an attempt at optimal prediction.  One thinks of the variables used to predict outcome simply as predictors that, in the past, would have helped one to make a better guess about what actually happened.  The predictors may have direct effects themselves or be correlated with some other unseen causal factor. However, those notions of direct and unseen cause are irrelevant to the method.  

Third, path analysis takes an intermediate approach.  One chooses the predictors for a model as a hypothesis about cause.  Rather than aim for optimal prediction, one aims for a set of variables that consistently describe the observed patterns of variation.  The quality of the causal interpretation is primarily evaluated by the consistency of the hypothesized pathways in capturing the observed variance in outcome.  Consistency roughly means relative stability in the magnitude of a pathway's effect under different circumstances.  Although that interpretation potentially offers some insight into cause and effect, the analytical method remains multiple regression.  One simply emphasizes the quality of a model as a potential causal interpretation rather than as an attempt at optimal prediction.

Consider a model in which we use genes as predictors of phenotype.  In a breeding program to improve yield, we want to predict offspring phenotype in order to make the best choice of breeding design.  Causality is irrelevant, we aim only for a good outcome.  By contrast, in a theoretical analysis of adaptation by natural selection, we want to understand the causal processes.  How do the genes that affect phenotype combine to determine morphology or behavior?  How does selection influence the underlying genes and the resulting phenotypic design in relation to performance?   We are after an understanding of the process.  The quality of prediction will, of course, be the primary way to interpret the causal model.  But a good prediction\break
    
\noterule
\end{minipage}
\end{figure}
\boxlabel{cause}

\begin{figure}[H]
\begin{minipage}{\hsize}
\parindent=15pt
\noterule
{\bf \noindent Box 5 --- continued\vskip2pt}
\noterule
\noindent  arising from the wrong underlying causal model is what we most want to avoid.  Prediction becomes a method for evaluation rather than the goal.

This article analyzes natural selection in relation to causal interpretations.  For that reason, I think of my models of multiple regression as models of path analysis.  In a different context, the same models could be thought of strictly as analyses of regression and prediction.      
\noterule
\end{minipage}
\end{figure}

\noindent genetics \autocite{falconer96introduction}.  Instead, I focus on general equations for selection and the transmission of information.  In my expressions, any predictors can be used including, but not limited to, all of the traditional genetic forms.  

Why bother with such abstractions?  Because many extensions to basic genetic theory have been developed to cope with nongenetic effects or to analyze selection independently of genetics \autocite{lynch98genetics}.  The literature tends to deal with each particular problem as a novel challenge that requires special theory. For example, maternal effects, kin selection, cultural evolution and institutional evolution in economics all have their distinct literatures and ways of framing problems.  Yet all of those problems are just examples of a general theory of selection and transmission.  In any particular application, the key is to express the causes of phenotypes (characteristics) and the causes of fitness (success) by a model, or hypothesis, of how various predictors combine to determine outcome.  A general theory expressed in terms of any choice of predictors defines the unifying conceptual framework \autocite{frank97the-price,frank98foundations}.

\subsection*{Fisher's average effect}

We can separate phenotype into components by 
\begin{equation*}
  z_i = \sum_j b_j x_{ij} + \Gd_i.
\end{equation*}
Each type, $i$, has $n$ different associated $x_i$ values, $x_{i1}, x_{i2},\ldots,x_{in}$.  From the perspective of multiple regression, the $x$'s are predictors, or independent variables, with respect to the phenotype, $z$.   Each $b_j$ is a partial regression coefficient of $z$ on $x_j$.  Roughly speaking, a partial regression coefficient, $b_j$, describes the average change in phenotype, $z$, for a change in the associated predictor variable, $x_j$.

We often focus on the general relation of a phenotype, $z$, to its components, $x_j$, rather than on the particular phenotype, $z_i$, of a particular type, $i$, in relation to its particular components, $x_{ij}$.  Thus, we may express the general relation between a phenotype and its components as
\begin{equation*}
  z = \sum_j b_j x_j + \Gd,
\end{equation*}
in which one understands that the particular values of $z$, $x_j$, and $\Gd$ vary for the different types, $i$, whereas the average effect of a predictor, $b_j$, is a property of the population.

The regression expression applies to any predictors, $x_j$.  We could use temperature, neighbors' behavior, another phenotype, epistatic interactions given as the product of allelic values, symbiont characters or an individual's own genes.  Fisher first presented this regression for phenotype in terms of alleles.  Suppose each $x_j$ is the presence or absence of an allelic type.  Then each $b_j$ describes the average contribution to phenotype for adding or subtracting the associated allelic type, and $b_j$ is called the average effect \autocite{fisher30the-genetical,crow70an-introduction,falconer96introduction}.  

Predicted phenotype is
\begin{equation}\label{eq:gDef}
  g = \sum_j b_j x_j.
\end{equation}
In genetic contexts, $g$ is often called the breeding value \autocite{falconer96introduction}.  Using $g$, we can partition phenotype into a predicted component and a residual component
\begin{equation}\label{eq:zReg}
  z = g + \Gd,
\end{equation}
where $\Gd = z-g$ is the difference between the actual value and the predicted value.  If we take the average of both sides, we get $\zbar = \gbar$, because $\bar{\Gd}=0$.

\subsection*{The components of heritability}

\subsubsection*{The part of phenotype not transmitted}

Typically, we only follow the transmission of the predictors.  For example, we may follow transmission of genes plus any other variables we choose.  Those effects that we include explicitly end up as part of the predicted phenotype, $g$, and as candidates for the transmitted phenotype.  All effects on phenotype not explicitly included as predictors end up in the residual, $\Gd$.  The split between the predicted phenotype and the residual is arbitrary.  If we add a new predictor, any additional effect of that predictor moves from the residual, $\Gd$, to the predicted phenotype, $g$.  Usually, we wish to give the best description of the causes of phenotype that we can.  Thus, our choice of predictors defines our hypothesis about the causes of phenotype, in the sense of path analysis discussed in \Boxx{cause}.

The part of phenotype associated with the particular set of predictors, $g$, defines one component of heritability. Aspects of phenotype not associated with the particular predictors in our model appear as a nontransmitted component of phenotype, $\Gd$, reducing the similarity of phenotype between ancestors and descendants associated with the predictors.  

\begin{figure}[H]
\begin{minipage}{\hsize}
\parindent=15pt
\noterule
{\bf \noindent\BoxLabel. Nonlinearity}
\noterule
\noindent Regression and path analysis are sometimes thought to be limited to linear and additive effects.  However, that is misleading.  Consider $z = bx + \Gd$.  Here, $b$ is the linear relation between $x$ and $z$.  However, it may be that $x=y^2$, in which the true underlying cause is $y$.  Thus, we are actually regressing on a nonlinear function of a causal variable, $y$.  Or, it may be that we start with $z = b_1x_1 + b_2x_2 + b_3x_3 + \Gd$.  This appears to be an additive model.   However, the underlying cause may be $x_1 = y_1$, and $x_2 = y_2$, and $x_3 = y_1*y_2$.  Thus, our model expresses nonlinearity and nonadditivity in the causes, $y$. 

In general, any nonlinear relation can be expressed by an additive sum of terms, in which the individual terms may be nonlinear.  Thus, regression can fully account for any nonlinearity by an additive sum of terms.  In practice, limitations arise because we may not know the correct nonlinear relation, and so cannot express the proper sum of nonlinear terms.  However, that is not a limitation of regression, but rather a limitation that arises from our ignorance.  Another method of analysis does not solve the problem of our ignorance. The point is that one must distinguish limitations arising from method from limitations arising from ignorance.  Confusing those different limitations is a common mistake.    
\noterule
\end{minipage}
\end{figure}
\boxlabel{nonlinearity}

\subsubsection*{Change in transmitted components of phenotype}

A second component of heritability arises from the stability of the effects associated with the predictors.  If a predictor has effect $bx$ in the original population and effect $b'x'$ in the second population, then the transmission of that predictor is associated with a change in phenotype $\GD (bx) = b'x'-bx$.  \Boxx{price} shows that we can express this change as
\begin{equation*}
  \GD (bx) = \left(\GD b\right) x + b'\left(\GD x\right).
\end{equation*}
Summing over the $j$ different predictors and using the definition of $g$ from \Eq{gDef} yields
\begin{equation}\label{eq:dG}
  \GD g =\sum \left(\GD b_j\right) x_{j} + \sum b_j'\left(\GD x_{j}\right).
\end{equation}
On the right side, the first term describes the change in the predicted value of a type that arises from the changes in the average effects of the predictors, $\GD b_j$, holding constant the predictor values, $x_j$.  For example, the average effect of an allele on phenotype may be frequency dependent.  Thus the average effect will change over time as the frequency of the allele changes in the population. The second term describes the change in the transmitted predictor values, $\GD x_j$, evaluated in the context of the average effects from the second population, $b_j'$.  For example, an allele may mutate into another form, thus weighting the average effect by a different amount.

The smaller the $\GD b$ and $\GD x$ values, the less the phenotype changes with respect to the transmitted predictors, and the higher the heritability associated with those predictors.  Equivalently, the more stable the predictors and their average effects, the greater the fidelity at which those particular predictors transmit the information accumulated by selection to the new population. 

The change in the predictors, $\GD x$, includes mutation as well as any other process that alters predictor values \autocite{frank95george,frank97the-price,frank98foundations,price95the-nature}.  For example, predictors in a descendant may derive from multiple ancestors. We can think of the mixing of predictors by considering the change in predictor values when derived from different sources.  In some cases, we may wish to alter the assignment of descendants to ancestors.  For example, a behavior may influence the frequency of nondescendant types.  To associate the behavioral phenotype with the change in frequency, we could assign those nondescendants to the ancestral behavior responsible for their presence \autocite{hamilton70selfish}.  In general, we can make such assignments in any way that we choose.  The key is that assigning different descendants to an ancestor may alter the change in predictor values between a descendant and its assigned ancestor.  Such changes may alter the fidelity at which information is transmitted \autocite{frank98foundations}.  I will take up that topic in the next article \autocite{frank13naturalvii}.

\subsubsection*{The part transmitted and the change during transmission}

The full, exact expression from \Eq{price} for the total evolutionary change is
\begin{equation*}
  \GD\zbar = \sum \GD q_i z_i + \sum q_i'\GD z_i.
\end{equation*}
We can partition phenotype as $z = g + \Gd$, the split between the part explained by the predictors of phenotype, $g$, and the part that is not explained by the set of predictors in our model for phenotype, $\Gd$.  From \Eq{zReg}, $\GD\zbar = \GD\gbar$ because $\bar{\Gd}=0$, thus
\begin{equation*}
  \GD\zbar = \GD\gbar = \sum \GD q_i g_i + \sum q_i'\GD g_i.
\end{equation*}
With $g_i = z_i - \Gd_i$, we get
\begin{equation}\label{eq:hPartition}
  \GD\zbar = \sum \GD q_i z_i -\sum \GD q_i \Gd_i + \sum q_i'\GD g_i.
\end{equation}
We can express each of these terms with a particular notation that emphasizes its interpretation
\begin{equation}\label{eq:totalChange}
  \GD\zbar = \GDs\zbar - \GDn\zbar + \GDt\zbar.
\end{equation}
On the right side, the terms are the change caused by selection, the change caused by the part of phenotype that is not associated with a transmitted predictor, and the change in the effects of the predictors during transmission.

\section*{Heritability and information}

This section focuses on the amount of information that populations accumulate by selection, and the various pro-\break  

\begin{figure}[H]
\begin{minipage}{\hsize}
\parindent=15pt
\noterule
{\bf \noindent\BoxLabel. Brief history of evolutionary partitions}
\noterule
\noindent \textcite{fisher18the-correlation,fisher30the-genetical} partitioned phenotype into its various genetic causes.  Quantitative genetics extended the partitioning of phenotype by genetic and nongenetic causes \autocite{falconer96introduction,lynch98genetics}.  Models of cultural evolution use culturally transmissible attributes as predictors of phenotype \autocite{dawkins76the-selfish,cavalli-sforza81cultural,boyd85culture}.  

Quantitative genetic models may also consider partitions of fitness into component causes.  Recent work on partitions of fitness was stimulated by \textcite{lande83the-measurement}.  Many subsequent studies expanded that approach, including various explicit descriptions based on path analysis \autocite{heisler87a-method,crespi89a-path-analytic,crespi90measuring,kingsolver91path,scheiner00using}.  I unified the different lines of study on partitions of phenotype and partitions of fitness \autocite{frank97the-price,frank98foundations}, motivated initially by Queller's quantitative genetic models of kin selection \autocite{queller92a-general,queller92quantitative}.

In the text, I mentioned that $rB-C>0$ can sometimes be interpreted in terms of group selection.  For example, if neighbors' phenotype, $y$, is an average character value in a local group, then $r$ can be defined as the regression of individual character value on group character value.  That group regression can be considered in a path analysis model, which is roughly the way in which \textcite{heisler87a-method} analyzed group selection.  In their article, they emphasized ``contextual analysis'' similarly to the way in which I have emphasized ``path analysis''.  \textcite{frank95mutual} and \textcite{taylor96how-to-make} also calculated $r$ by regressing group value on individual value in several models, following a long tradition that blurred the mathematical distinction between kin and group selection \autocite{hamilton75innate,frank86hierarchical}.   

Some of the multivariate analyses of fitness attempt to predict evolutionary dynamics, and therefore must make explicit assumptions about the distribution of phenotypes and the nature of heritability.  I do not discuss dynamics; my models do not require any of those extra assumptions.      
\noterule
\end{minipage}
\end{figure}
\boxlabel{partition}

\noindent cesses that degrade or alter the transmission of that information.  Some of the forms given here include the classic genetic measures of heritability as special cases.  However, I do not emphasize those connections.  Rather, I focus on general expressions given in terms of the full Price equation for total evolutionary change and based on predictors that may be chosen in any way.  Different problems and goals will lead one to choose different sets of predictors or underlying causal schemes for phenotypes.  The results here apply to any choice of predictors and causal scheme. 

We start with \Eq{hPartition}, the partition of phenotypic change into components 
\begin{equation*}
  \GD\zbar = \sum \GD q_i z_i -\sum \GD q_i \Gd_i  + \sum q_i'\GD g_i .
\end{equation*}
The first term on the right side is the selection component, $\GDs\zbar$. From \Eq{alphaJ}, $\GDs\zbar=\Ga_z\J$, where $\Ga_z$ changes scale between phenotype, $z$, and the gain in information by selection, $\J$. Thus, 
\begin{equation*}
  \GD\zbar = \Ga_z\J  -\sum \GD q_i \Gd_i  + \sum q_i'\GD g_i.
\end{equation*}
Here, selection happens in the initial (parental) population, causing a gain in information, $\J$. On the phenotypic scale, that gain in information is $\Ga_z\J$.  The remaining terms include processes that cause loss of information during transmission or cause other changes to phenotype.

\subsection*{The part of phenotype not transmitted}

Start by assuming that the predictors and their effects do not change during transmission, $\GD g_i=0$.  That assumption reduces total change to
\begin{equation*}
  \GDh\zbar = \Ga_z\J  -\sum \GD q_i \Gd_i
\end{equation*}
where $\GDh=\GDs-\GDn$ denotes the heritable component of selection, which is the total selection, $\GDs$, minus the part of selective change that is not associated with predictors, $\GDn$.  The part not associated with predictors is not explicitly transmitted within the given model of phenotype.  

The second term, $\sum \GD q_i \Gd_i$, has the general form (\Eq{alphaJ}) of the change in information
\begin{equation*}
  \sum \GD q_i z_i = \Ga_z\J,
\end{equation*}
which holds for any choice of $z$.  Thus, letting $z\equiv\Gd$, we obtain $\sum \GD q_i \Gd_i=\Ga_\Gd\J$.  Putting this into the original expression yields
\begin{equation*}
  \GDh\zbar = \Ga_z\J  -\Ga_\Gd\J = (\Ga_z  -\Ga_\Gd)\J.
\end{equation*}
The scale change terms, $\Ga$, have the important additivity property that, in general, $\Ga_a+\Ga_b=\Ga_{a+b}$.  Thus, 
\begin{equation*}
  \Ga_z  -\Ga_\Gd = \Ga_{z-\Gd} = \Ga_g,
\end{equation*}  
because $g=z-\Gd$.  The expression for the change in phenotype, ignoring the change during transmission in the predictors and their effects, is
\begin{equation}\label{eq:gJ}
  \GDh\zbar = \Ga_g\J.
\end{equation}
This expression is the information gain by selection, $\J$, scaled by $\Ga_g$, which relates the predicted phenotype, $g$, to the information accumulated by selection.  Because $g=z-\Gd$, we see that the amount of information transmitted is degraded by $\Gd$, the fraction of the phenotype, $z$, that is not explained by the predictors.  

\subsection*{Change in transmitted components of phenotype}

When we add back the remaining term to \Eq{gJ}, we obtain the full expression for phenotypic change as
\begin{equation*}
  \GD\zbar = \Ga_g\J + \sum q_i'\GD g_i.
\end{equation*}
The last term is the change in the transmitted components of phenotype.  From \Eq{dG}, those components include changes in the predictors and changes in the effects of the predictors.  A predictor's effect is its associated multiple regression coefficient.  Multiple regression coefficients often change with context. On the one hand, the true underlying causal effect may change. On the other hand, our model of causality may not be exactly right, in which case shifting context will cause the assigned role of different predictors to change, even though the underlying causal effects of those predictors may not have changed.  

Various approaches may be taken to evaluate the accuracy of the causal model, such as the stability of the predictor effects under changing context \autocite{li75path}.  Typically, a better causal model has predictors with greater stability, shifting the components of total change more strongly to the $\Ga_g\J$ information term.  That increase in the information term is usually advantageous with respect to interpretation, because it is often hard to evaluate the meaning of changes in predictors and their effects in the second term.  

Suppose, for example, that a significant component of phenotype is not explained by a stable set of predictors.  Is the information accumulated by selection in the initial population lost during transmission because it is not associated with any transmissible component?  Or, is that information transmitted by other predictors that are not included in our model? If the information does transmit by predictors not in our model, that information contributes to the second term with changing values of the predictors and their effects. Such changes are hard to interpret, because many different processes can potentially alter the predictors and their effects.

These fundamental equations of selection and evolution are, in a way, rather arbitrary, because they depend so strongly on the particular set of predictors that one chooses.  What can we conclude?  First, the equations are always true, and so give us a clear sense of the essential nature of selection, information and evolution.  Second, a key part of understanding any problem concerns choosing the right set of predictors.  Third, simple genetic models provide a good starting point in many cases, but rarely define a complete set of predictors and an accurate expression of causality.  If one is able to model the causal scheme well, the analysis will often be simple and natural.  I have emphasized a path analysis interpretation for the regression expressions, because path analysis emphasizes the choice of a good causal model.  

\subsection*{Fisher's fundamental theorem}

If we hold the predictors and their effects constant, then using \Eq{gJ}, the change in mean log fitness is
\begin{equation*}
  \GDh\mbar = \Ga_g\J
\end{equation*}
for $m=g+\Gd$.  This expression for change in fitness, holding constant the predictors and their average effects, provides a generalization of Fisher's fundamental theorem of natural selection.  Fisher used the presence or absence of allelic types as predictors, and the associated value of predicted fitness, $g$, as the genic value of fitness.  With those definitions, the expression here is equivalent to Fisher's theorem.  To translate back to the particular notation that Fisher used, one would translate the definitions for $\Ga_g$ and $\J$ into Fisher's forms.  \textcite{frank97the-price} provides the tools for the translation, following \textcite{price72fishers} and \textcite{ewens89an-interpretation}.  The point here is that Fisher's theorem holds for any choice of predictors, as emphasized in \textcite{frank97the-price}.

\section*{Causes of fitness}

The expression $\GDs\zbar=\Ga_z\J$ associates the accumulation of information by selection, $\J$, with the selective component of phenotypic change.  But that expression does not tell us why the association occurs.  The phenotype may directly influence fitness. Alternatively, the phenotype may have no direct effect on fitness, but instead may be associated with some other process that influences fitness.  A significant part of evolutionary analysis concerns evaluating the causes of fitness (\Boxx{partition}).  

We may analyze the causes of fitness in the same way that we analyzed the causes of phenotype.  We write our model, or hypothesis, for the causes of fitness as the regression equation
\begin{equation}\label{eq:lande}
  w = \Gf + \pi z + \sum a_k y_k + \Ge.
\end{equation}
Here, $\Gf$ is the baseline fitness when all other terms are zero; $\pi$ is the average direct effect of the phenotype $z$ on fitness, holding constant the other predictors of fitness; and $a_k$ is the average effect of the other predictors of fitness, $y_k$.  We may use any number of other predictors, and those predictors may be defined in any way, including factors in the model for phenotype.  For example, predictors $y_k$ can be alleles, nonlinear interactions between combinations of alleles, symbionts, maternal effects, cultural or environmental attributes, other phenotypes, phenotypes of neighbors, and so on.  The residual, $\Ge$, is the difference between the predicted value of fitness for a given set of predictors and the actual fitness.  

\subsection*{A simple example}

To study the role of different predictors of fitness, it is useful to reduce the model to just the direct effect, $z$, and one indirect effect, $y$, yielding
\begin{equation*}
  w = \Gf + \pi z + a y + \Ge.
\end{equation*}
In this partial regression equation, it is helpful to write out the regression coefficients in full notation to emphasize their interpretation.   The partial regression coefficient $\pi = \preg{wz}{y}$ is the average effect of $z$ on $w$ holding $y$ constant, and $a=\preg{wy}{z}$ is the average effect of $y$ on $w$ holding $z$ constant, thus
\begin{equation}\label{eq:bivarEx}
  w = \Gf + \preg{wz}{y} z + \preg{wy}{z} y + \Ge.
\end{equation}

\subsection*{Condition for the increase of a phenotype by selection}

Using the standard covariance form for selection based on \Eq{fitnessReg}, the partial change in $z$ caused by selection is
\begin{equation*}
  \wbar\GDs\zbar = \cov(w,z),
\end{equation*}
which simply states that $z$ increases by selection when it is positively associated with fitness.  However, we now have the complication shown in \Eq{bivarEx} that fitness also depends on another predictor, $y$.  If we expand the covariance using the full expression for fitness in \Eq{bivarEx}, we obtain
\begin{equation*}
  \wbar\GDs\zbar = \preg{wz}{y} V_z + \preg{wy}{z} \cov(y,z).
\end{equation*}
If we replace the covariance term by the product of a regression coefficient and a variance, $\Gb_{yz}V_z$, we have
\begin{equation}\label{eq:dzHR}
  \GDs\zbar = \left(\preg{wz}{y} + \preg{wy}{z} \Gb_{yz}\right)V_z/\wbar.
\end{equation}
The condition for the increase of $z$ by selection is $\GDs\zbar > 0$.  The same condition using the terms on the right side is
\begin{equation}\label{eq:HRregr}
  \Gb_{yz}\preg{wy}{z}  + \preg{wz}{y}  > 0.
\end{equation}
Let us use an abbreviated notation for the three terms\begin{align*}
  \Gb_{yz} &= r\\
  \preg{wy}{z} &= B \\
  \preg{wz}{y} &= -C.
\end{align*}
The first term, $\Gb_{yz}=r$, describes the association between the phenotype, $z$, and the other predictor, $y$. An increase in $z$ by the amount $\GD z$ corresponds to an average increase of $y$ by the amount (see \Boxx{scale})
\begin{equation*}
  \GD y = r\GD z.
\end{equation*}
The second term, $\preg{wy}{z}=B$, describes the direct effect of the other predictor, $y$, on fitness, holding constant the focal phenotype, $z$.  The third term, $\preg{wz}{y}=-C$, describes the direct effect of the phenotype, $z$, on fitness, $w$, holding constant the effect of the other predictor, $y$.  

Using the abbreviated notation, the condition for the increase in $z$ by selection is
\begin{equation*}
  rB - C > 0.
\end{equation*}
The following sections interpret this condition in terms of three different biological scenarios.

\begin{figure*}[t]
\includegraphics[width=0.75\hsize]{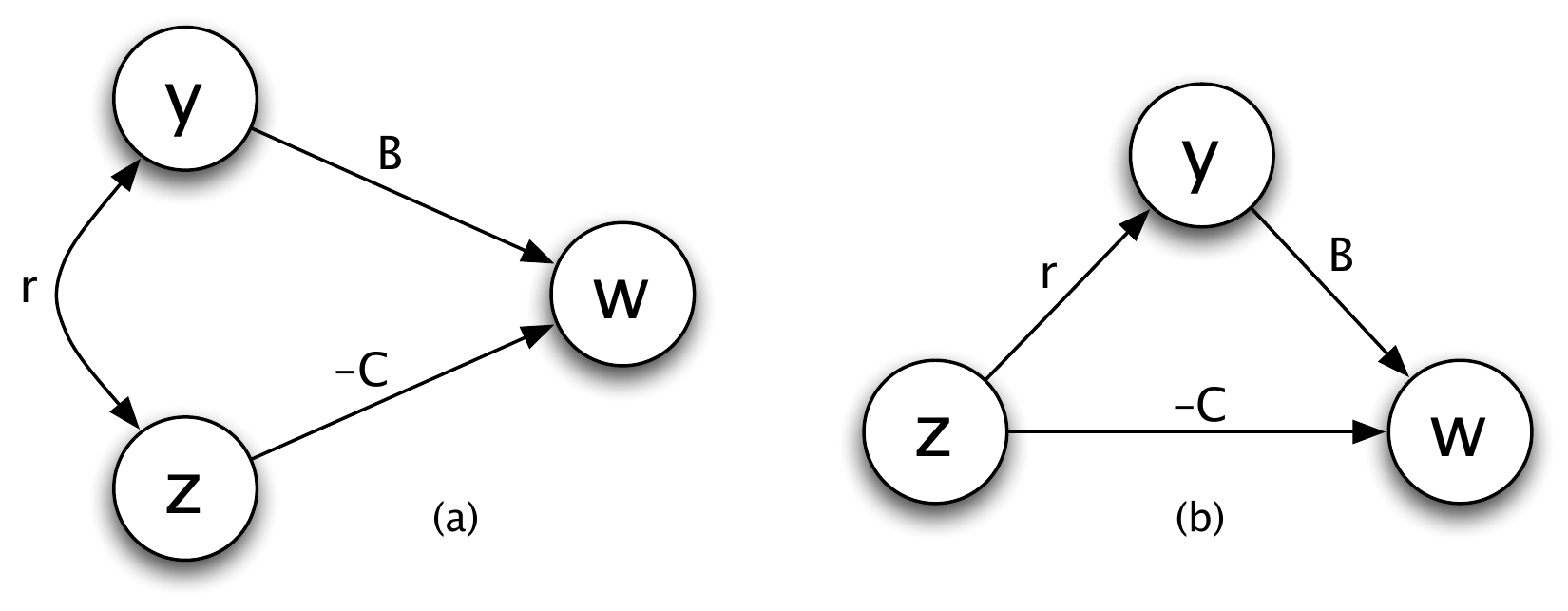}
\caption{Path diagrams for the effects of phenotype, $z$, and secondary predictor, $y$, on fitness, $w$.  (a) An unknown cause associates $y$ and $z$. The arrow connecting those factors points both ways, indicating no particular directionality in the hypothesized causal scheme.  (b) The phenotype, $z$, directly affects the other predictor, $y$, which in turn affects fitness.  The arrow pointing from $z$ to $y$ indicates the hypothesized direction of causality. }
\label{fig:path}
\end{figure*}

\subsection*{Interactions between two species}

I trace the effects of phenotype $z$ in species A and phenotype $y$ in species B on the fitness of types from species A \autocite{frank94genetics,frank95the-origin,frank97models}.  One may think of species B as an ecological partner that can influence the fitness of types from species A.  Here, fitness always refers to effects on species A.  

\subsubsection*{Unknown cause of association}

I follow the path diagram in Fig.~1a.  Increases in the phenotype, $z$, by an amount $\GD z$, reduce fitness by $-C\GD z$.  Increases in the phenotype $y$ directly benefit fitness by $B\GD y$.  The $z$ and $y$ phenotypes are associated by $r$, although no specific cause is known.  It may be that similar phenotypes tend to settle in the same area, or that a common environment of temperature and nutrients causes a phenotypic association.  In any case, as $z$ increases, the associated value of $y$ changes on average by $\GD y = r\GD z$ and, equivalently, $B\GD y=rB\GD z$. 

Tracing the pathways in Fig.~1a, an increase in the direct phenotype by $\GD z$ causes a change in fitness by $(rB-C)\GD z$, which is greater than zero when $rB-C>0$.  Thus, selection may favor an increase in $z$ even though $z$ directly decreases fitness, because the benefit from species B's phenotype, $y$, in proportion to $rB$, may outweigh the direct cost, $-C$.  

\subsubsection*{Direct cause of association}

Alternatively, suppose that the phenotype $z$ directly enhances the vigor of its partners from species B. That direct effect of $z$ on species B causes an increase in the benefit, $y$, that species B provides back to those with phenotype $z$.  Fig.~1b shows this direct cause of $y$ by $z$.  The condition for $z$ to be positively associated with fitness and to increase by selection remains $rB-C>0$.  However, the interpretation differs.  In this case, $z$ directly influences its neighbors' phenotype, $y$, rather than being associated with $y$ by some unknown cause.

\subsection*{Body temperature}

Suppose $z$ is body temperature, which imposes a direct effect $-Cz$ on fitness.  That direct cost may arise because body temperature raises the rate at which energy is used. Let $y$ be speed of response to a challenge, such as a predator attack.  Faster response provides a direct benefit, $By$.  An unknown cause may associate temperature, $z$, and response rate, $y$, by an amount $r$ (Fig.~1a).  For example, sunshine may directly raise temperature and simultaneously increase response to attack by providing better visual opportunities.  Alternatively, temperature, $z$, may directly raise response rate, $y$, by increasing the responsiveness of muscles (Fig.~1b).  In either case, selection favors an increase in body temperature if $rB - C > 0$.  

\subsection*{Social evolution and group selection}

The phenotype $z$ may be a costly altruistic behavior that helps neighboring individuals \autocite{hamilton70selfish,queller92a-general,queller92quantitative,frank98foundations}.  The direct effect on fitness is $-Cz$.   Neighbors have phenotype $y$ that provides a benefit, $By$, back to the original individual.  An association, $r$, between $z$ and $y$ may arise in a variety of ways.  

Some unknown cause may associate $z$ and $y$ (Fig.~1a).  For example, shared cultural, environmental or genetic variation may cause related behavior.  Or a shared symbiont may cause an association.  In general, any association in the predictors of phenotype will cause an association of phenotypic values.  

In other cases, the altruistic phenotype, $z$, may directly enhance neighbors' beneficial behavior, $y$, in proportion to $r$ (Fig.~1b).  For example, the level of $y$ in the neighbors may depend on the probability of the neighbors' survival.  If an increase in $z$ raises neighbors' survival in proportion to $r$, that increase in survival enhances the expression of the neighbors' behavior, $y$, which has a beneficial effect on fitness of $By$.  

Whether $r$ arises from unknown causes (Fig.~1a) or from the direct effect of $z$ on $y$ (Fig.~1b), we can trace the effect of an increase in $z$ on fitness.  The condition for an increase in $z$ to raise fitness is $rB-C>0$. 

In some cases, we may interpret the condition $rB-C>0$ in terms of group selection \autocite{hamilton75innate}.  For example, $z$ may measure individual restraint in the harvesting of nonrenewable resources \autocite{frank95mutual}.  Greater restraint reduces the direct benefit to the individual, because it means less resource harvested, with an effect on fitness of $-Cz$.  Neighbors' phenotype, $y$, may be the average restraint among individuals in a local group with regard to harvesting nonrenewable resources.  

Greater group restraint provides a benefit to all members of the group, including our focal individual, by providing greater local productivity through maintenance of nonrenewable resources.  The benefit of group restraint on individual fitness is $By$.  The association between an individual's phenotype, $z$, and the group phenotype, $y$, is $r$.  Thus, when $rB-C > 0$, individual restraint evolves and provides a joint benefit to all group members.  Here, the two predictors of fitness are individual behavior, $z$, and average group behavior, $y$.  This type of group selection is just a special case of partitioning the causes of fitness, in which one of the predictors is a group attribute (\Boxx{partition}).  

\section*{Causal structure}

All of these examples share a common causal structure.  We are interested in the change in a phenotype, $z$, caused by selection.  Fitness depends on two predictors: the phenotype of interest, $z$, and another predictor, $y$.  In all cases, the condition for the increase in $z$ by selection is $rB-C>0$.  This condition is just the partition of the causes of fitness into two components.  The direct effect on fitness of $z$ is $-C$, and the direct effect of $y$ is $B$.  We multiply $y$ by $r$ to change the scale of the effect from $y$ to $z$, because the net effect must be the relation between $z$ and fitness, $w$.  

We can see the logical relations and the units for the various scales by writing out the full notation
\begin{equation}\label{eq:hrEquiv}
  rB - C = \Gb_{yz}\preg{wy}{z}  + \preg{wz}{y}.
\end{equation}
\Boxx{regression} shows that a regression coefficient, $\Gb_{xy}$, has units $\GD x/\GD y$.  Taking the terms of the above equation in order from left to right, the units are
\begin{equation}\label{eq:hrUnits}
  \Gb_{yz}\preg{wy}{z}  + \preg{wz}{y} \equiv \frac{\GD y}{\GD z} \frac{\GD w}{\GD y} + \frac{\GD w}{\GD z}
  		\equiv \frac{\GD w}{\GD z}.
\end{equation}
The ratio $\GD w/\GD z$ is the change in fitness, $w$, per unit change in the phenotype, $z$.  That ratio is the slope of fitness on phenotype.  When the slope is positive, selection favors the increase of the phenotype.  In any analysis of this sort, the term
\begin{equation}\label{eq:rDef}
  r = \Gb_{yz} = \frac{\GD y}{\GD z}
\end{equation} 
rescales changes of the secondary predictor, $\GD y$, with respect to changes in the primary scale, $\GD z$.  

The key point is that $rB-C>0$ simply partitions fitness into the direct effect of a phenotype plus the indirect effect through a secondary predictor.  The true causal structure will, of course, frequently depend on multiple secondary causes, as in \Eq{lande}. Multiple causes lead to an expanded expression for the increase of $z$ caused by selection, $\GDs\zbar$, as 
\begin{equation*}
  \sum r_iB_i - C > 0,
\end{equation*}
in which each $r_i$ is the regression of $y_i$ on $z$, and each $B_i$ is the partial regression of $w$ on $y_i$ holding constant the other factors.  One may also need to consider cascading causes or hidden factors in the sense of path analysis \autocite{li75path}. The simple expression $rB-C>0$ should be thought of as a convenient example to illustrate the logic of partitioning the causes of fitness, or as the expression of simplified models that isolate two opposing processes.  

In this section, I have analyzed the partitioning of fitness.  I have not discussed the partition of phenotype into components, $z = g + \Gd$, where $g$ is the sum of the predictors of phenotype.  The amount of information accumulated by selection that can be transmitted depends on the slope of fitness, $w$, relative to the transmissible predictors of phenotype, $g$.  If we think of $g$ in terms of the genetic predictors of phenotype, then $r$ can be interpreted as a genetic relatedness coefficient, and $rB-C>0$ calls to mind Hamilton's rule from the theory of kin selection \autocite{hamilton70selfish}.  The next article takes up the relations between kin selection and the general analysis of the causes of fitness and the causes of phenotype \autocite{frank13naturalvii}. A full evolutionary analysis also requires attention to other causes of change, $\GDt\zbar$, in \Eq{totalChange} \autocite{frank97the-price,frank98foundations}.

It is important to relate the causes of fitness to information, which is the ultimate scale for selection. \Boxx{infoFit} connects the partitions of fitness in this section to the expressions of information given earlier in this article. 

\begin{figure}[H]
\begin{minipage}{\hsize}
\parindent=15pt
\noterule
{\bf \noindent\BoxLabel. Information and the causes of fitness}
\noterule
\noindent Changes caused by selection can always be related to the change in information accumulated by the population.  For example, the change in phenotype caused by selection from \Eq{alphaJ} is
\begin{equation*}
  \GDs\zbar = \Ga_z\J,
\end{equation*}
where $\J$ is the change in information by selection, and $\Ga_z$ relates the scale of information to the scale of phenotype.  We can examine the units of the scaling term 
\begin{equation*}
  \Ga_z = \frac{\Gb_{zw}}{\Gb_{mw}},
\end{equation*}
which is the ratio of two regression coefficients (\Eq{alphaDef}).  A regression coefficient, $\Gb_{zy}$, has units $\GD z/\GD y$, when used as a scaling relation for changes in average values (\Boxx{regression}).  Thus, the units for the scaling relation, $\Ga_z$, are
\begin{equation*}
  \frac{\Gb_{zw}}{\Gb_{mw}} \equiv \frac{\GD z}{\GD w} \frac{\GD w}{\GD m} = \frac{\GD z}{\GD m}.
\end{equation*}
The term $\GD m$ has units of change in log fitness.  Changes in log fitness are equivalent to changes in information, $\J$ (\Eq{mJ}).  To emphasize that $\J$ is a change in information, write the units on $\J$ as $\GD I$.  Thus, the scaling factor 
\begin{equation*}
  \Ga_z \equiv \frac{\GD z}{\GD I}
\end{equation*}
is the change in phenotype relative to the change in information.

One must learn to read the regression coefficients as scaling factors that change units.  Once one learns to recognize the scale changes, and the key units such as information and phenotype, the fundamental equations can be read like a sentence.  When analyzing selection, I prefer information as the ultimate scale, because selection is the process by which populations accumulate information.  

With that background, I present a long sentence to translate the causes of fitness into an expression for the change in information.  Start with \Eq{dzHR} and divide both sizes by $\Ga_z$, yielding
\begin{equation*}
   \GDs\mbar = \frac{\GDs\zbar}{\Ga_z} = \Gb_{mw}
   										\left(\frac{\preg{wz}{y} + \preg{wy}{z} \Gb_{yz}}{\Gb_{zw}}\right)\frac{V_z}{\wbar}.
\end{equation*}
The units are
\begin{equation*}
   \frac{\GDs\zbar}{\Ga_z} \equiv \GD z \left(\frac{\GD I}{\GD z}\right) = \GD I,
\end{equation*}
the change in information by selection.  All of the regression coefficients in the prior equation change scales for the various terms, and we also have $V_z/\wbar$, which has units $\GD z^2/\GD w$.   The net units of the long right side are $\GD I$, the change in information.  The right side appears complex.  But each term has a simple, readable meaning with respect to the effect of a predictor on fitness, and the scale changes required to transform those effects into the common units of information.  To understand selection, we often need to decompose fitness and phenotypes into their component causes.  Such decomposition requires that we combine all the components properly to recover the correct scale of analysis.    
\noterule
\end{minipage}
\end{figure}
\boxlabel{infoFit}

\section*{Discussion}

I first partitioned phenotype with respect to a set of hypothesized causes. I then partitioned fitness with respect to a different set of hypothesized causes.  Finally, I placed those partitions of phenotype and fitness into a general expression for selection and evolutionary change.  Those steps allowed me to express heritability, selection and evolutionary change in terms of causal components.  

I also translated the standard expressions of selection and evolution, given in terms of regressions, covariances and variances, into expressions for the change in information.  In my view, selection is best interpreted as the accumulation of information by populations \autocite{frank12naturalb}.  Other evolutionary processes often cause a decay in the transmission of information.  The information expressions allow one to read the equations of selection and evolution as if they were sentences.  Those sentences express the fundamental relations between the causes of phenotypes and fitness and the consequences for the change in information by evolutionary processes.

I showed that the commonly used regressions coefficients in models of selection and evolution can be understood as coefficients for the change in scale with respect to the ultimate scale of information (\Boxx{scale}).  For example, the change in a phenotype caused by selection can be understood as a rescaling of the change in information accumulated by selection.  Certain measures of heritability, often expressed as regression coefficients, are the change in the scaling of information from one phenotype to another. For example, a parent-offspring regression may describe the change in scale between parent and offspring phenotype with respect to the underlying information content in those phenotypes.  

My extended development in terms of causal components and information may, at first, seem like a lot of technical complication.  We are, after all, simply modeling selection, heritability and other widely studied evolutionary processes.  Many models of those processes seem more direct and concise. My goal is to go beyond common calculations or common applications.  The more abstract and exact models here provide a conceptual guide for understanding how selection actually works, how populations accumulate information, and how that information is transmitted or lost.  

I have also traded the certainty of the standard models of genetics for the uncertainty that arises when we freely choose our predictors as causal hypotheses.  In my view, the apparent certainty of genetics is often misleading.  We know that many factors influence phenotypes in addition to the narrowly defined allelic types of genes.  Traditionally, a specific extended model deals with each additional factor: cytoplasmic inheritance, nonlinear genetic interactions, maternal effects, social interactions, and so on.  By describing each of those aspects as a special situation, one ends up with a catalog of special models.  

The models here show how to think in general about a variety of causal structures.  Those models are only as good as the particular hypothesized system of causality that we choose.  But that is also true for genetic models and for every other model, whether or not we admit it openly.  Here, I have traded the false sense that there are a few standard models for the more realistic view that one has to bring a good hypothesis to an analysis in order to get a good understanding of phenotypes and selection.

\textcite{hamilton70selfish} made clear the central role of causal analysis in kin selection theory
\begin{quote}
Considerations of genetical kinship can give a statistical reassociation of the [fitness] effects with the individuals that cause them.
\end{quote} 
The seemingly endless debates about kin selection arise from failure to recognize that the theory is ultimately a way of framing causal hypotheses \autocite{frank97the-price,frank98foundations}.  The following article develops kin selection as a method of causal modeling.

\section*{Acknowledgments}

National Science Foundation grant EF-0822399 supports my research.  

\vskip.6in

\bibliography{main}

\end{document}